\documentclass[twocolumn,superscriptaddress,nofootinbib,floatfix]{revtex4-2}
\usepackage{graphicx} % Required for inserting images
\usepackage{amsmath} 
\usepackage{tabularx}
\usepackage{color} % for comments
\usepackage{array} % for fixing table widths
\newcommand\Tstrut{\rule{0pt}{2.6ex}}         % `top' strut for tables
\newcommand\Bstrut{\rule[-0.9ex]{0pt}{0pt}}   % `bottom' strut for tables

\begin{document}
\title{Tunable-fidelity wave functions for the \textit{ab initio} description of scattering and reactions}
\author{Konstantinos Kravvaris}
\email{kravvaris1@llnl.gov}
\affiliation{Lawrence Livermore National Laboratory, P.O. Box 808, L-414, Livermore, California 94551, USA}
\author{Sofia Quaglioni}
% \email{quaglioni1@llnl.gov}
\affiliation{Lawrence Livermore National Laboratory, P.O. Box 808, L-414, Livermore, California 94551, USA}
\author{Petr Navr\'atil}
% \email{quaglioni1@llnl.gov}
 \address{TRIUMF, 4004 Wesbrook Mall, Vancouver, British Columbia, V6T 2A3, Canada}% \date{July 2023}
% \author{Guillaume Hupin}
% %\email{hupin@ipno.in2p3.fr}
% \address{Universit\'e Paris-Saclay, CNRS/IN2P3, IJCLab, 91405 Orsay, France}
\begin{abstract}
\edef\oldrightskip{\the\rightskip}
\begin{description}
\rightskip\oldrightskip\relax
\item[Background] The no-core shell model (NCSM) is an \textit{ab initio} method that solves the nuclear many-body problem by expanding the many-particle wave function into a (typically) harmonic oscillator basis and minimizing the energy to obtain the expansion coefficients. Extensions of the NCSM, such as its coupling with microscopic-cluster basis states, further allow for an \textit{ab initio} treatment of light-ion nuclear reactions of interest for both astrophysics and nuclear technology applications. 
A downside of the method is the exponential scaling of the basis size with increasing number of nucleons and excitation quanta, which limits its applicability to mass $A\lesssim 16$ nuclei, except for variants where the basis is further down-selected via some truncation scheme. 
\item[Purpose] We consider a basis selection method for the NCSM that  captures the essential degrees of freedom of the nuclear wave function leading to a favorable complexity scaling for calculations and enabling \textit{ab initio} reaction calculations in $sd$-shell nuclei. 
\item[Methods] The particle configurations within the NCSM basis are ordered based on their contribution to the first moment of the Hamiltonian matrix that results from the projection onto the many-body basis. The truncation scheme then consists in retaining only the lowest-first-moment configurations, which typically contain only few many-body basis states (Slater determinants). As the energy threshold above which configurations are disregarded is increased, the size of the basis becomes an almost-continuous variable, allowing for tunable fidelity in the obtained wave functions.  
The resulting wave functions can then be used directly in \textit{ab initio} reaction calculations.
\item[Results] We present calculations for $^7$Li and $n+^{12}$C scattering using nucleon-nucleon interactions derived from chiral effective field theory and softened using the similarity renormalization group method. The obtained energy levels invariably demonstrate exponential convergence with the size of the basis, and we find improved convergence in scattering calculations. To demonstrate the possibilities enabled by the approach, we also present a first calculation for the scattering of neutrons from $^{24}$Mg. 
\item[Conclusions] The method presented in this work appears promising for future studies of nuclei with mass $A>16$, opening multiple future research directions impacting both nuclear astrophysics and nuclear technology applications. 
\end{description}
\end{abstract}
\maketitle

\section{Introduction}
The description of nuclear structure 
from first principles (or, \textit{ab initio}) plays a key role in modern-day nuclear theory, with various methods developed in recent years exhibiting remarkable successes~\cite{Hergert2016,Tichai2016,Hergert2020,Stroberg2021,Hu2022}. 
Many such methods are able to reach mid-mass and heavier nuclei as they exhibit polynomial scaling with  the number of nucleons in the system~\cite{Soma2011,Hagen2014,Hergert2016}. 
Alternatively, quasi-exact methods such as quantum Monte Carlo~\cite{Carlson2015} or the no-core shell model (NCSM)~\cite{Navratil2000,Barrett2013} do not enjoy such a  
favorable scaling, with an exponential growth in problem dimension being typical as heavier nuclei are targeted. Nevertheless, 
these methods come with increased precision, making 
them the state of the art in the systems where they can be applied. 

This trove of methods has been readily applied to the description of nuclear structure properties but their (direct) adaptation to 
the calculation of nuclear reactions has been slower and 
more limited: In light and medium-mass systems, most developments centered around the scattering of single nucleons from nuclei~\cite{Nollett2007,Hagen2012,Papadimitriou2013,Lynn2016, Shirokov2016, Rotureau2018, Flores2023} though some extensions to the treatment of reactions~\cite{Rotureau2020} have also been considered. 
Approaches utilizing lattice effective field theory~\cite{Elhatisari2015} have also been brought forward, targeting the scattering of $^4$He nuclei from $N=Z$ targets. 
The NCSM with continuum (NCSMC)~\cite{Baroni2013L,Baroni2013C} is perhaps the most advanced of this class of methods, having enabled a wide range of predictive \textit{ab initio} calculations of nucleon and deuteron-induced scattering~\cite{Navratil2016,Kumar2017,Kravvaris2020b,Holl2021,Hebborn2022}, capture~\cite{McCracken2021,Kravvaris2023}, and fusion reactions~\cite{Hupin2019} in light nuclei.  
%description of both scattering and reactions in a large range of cases (see Ref.~\cite{Navratil2016} for a recent review).
Another pathway to linking nuclear structure approaches to the description of scattering that has recently received significant attention lies in the extraction of effective nucleon-nucleus (optical) potentials from the properties of the nuclear wave function of the aggregate system~\cite{Vorabbi2016,Gennari2018,Idini2019,Burrows2019}. 
Such potentials, while useful for the description of high-energy scattering of single nucleons, do not provide accurate results in the region of resolved resonances that typically characterizes lower energies~\cite{Hebborn2023}.
Finally, we should also mention that
multiple \textit{ab initio} nuclear structure methods 
%obtaining accurate results for 
have been successfully extended to the description of photonuclear cross sections~\cite{Stetcu2007,Orlandini2014,Bacca2014} and electroweak response functions~\cite{Lovato2013,Lovato2023}.

The wave function ansatz of the NCSMC approach combines microscopic cluster states in the spirit of the resonating group method (RGM)~\cite{RGM,RGM3} that describe the relative motion of the reaction fragments with eigenstates of the aggregate system that capture the many-body correlations whose effects dominate at short distances. 
The many-body wave functions of the projectile, target and aggregate systems are all computed within the NCSM.     
Thus, the NCSMC faces some of the same computational limitations as the NCSM, particularly when applied to the description of reactions for nuclei heavier than oxygen. 
However, the basis states required to describe the wave functions of the reactants %in the NCSM 
are not all equally important when it comes to computing low-energy scattering and reactions. % in the NCSMC. 
While the 
 projectile's and target's energies must be described accurately to correctly reproduce reaction thresholds, high-fidelity details in their wave functions may not be as essential; it is therefore interesting to explore the use of tunable-fidelity NCSM wave functions within the NCSMC. 

Work along these lines was first carried out in Ref.~\cite{Navratil2010}, when the NCSM combined with the resonating group method (NCSM/RGM)~\cite{Quaglioni2008,Quaglioni2009} was applied to $n$+$^{12}$C and $n$+$^{16}$O scattering by leveraging the importance-truncated NCSM (IT-NCSM)~\cite{Roth2007,Roth2009}. 
Wave functions obtained in the IT-NCSM were later also used in NSCMC calculations of $n+^{16}$C in Ref.~\cite{Smalley2015}, as well as in the description of the properties of $^9$Be~\cite{Langhammer2015}. 

The many-body basis selection in the IT-NCSM is founded in many-body perturbation theory, where the action of the Hamiltonian on a starting vector will determine the basis states that will be retained. 
This approach can be considered part of a broader class of methods that use the Hamiltonian to dynamically select the relevant components of the basis, while remaining agnostic to the specifics of the nuclear interaction. 
Another approach is to statically pre-select the basis on which the many-body wave function will be expanded based on symmetry considerations. An example of this class of methods is the symmetry-adapted no-core shell model~\cite{Launey2020}, for which initial steps towards its extension to the description of nuclear scattering was presented in Ref.~\cite{Mercenne2022}. 
In this case, the selection of the basis relies on symmetry assumptions and the resulting calculations are significantly more complicated but fare remarkably well when considering the convergence of symmetry-dominated observables, such as collective E2 transition rates. 

%These two ``extremes" represent two different philosophies when it comes to basis truncation. From the hamiltonian-driven IT-NCSM, to the basis pre-selection described in~\cite{Launey2020,Mercenne2022}, each approach comes with the aforementioned distinct sets of advantages and disadvantages. 
Here, we implement in the context of the NCSM an intermediate approach first introduced in Ref.~\cite{Horoi1994} for the large-scale shell model~\cite{Brown1988,Caurier2005} (LSSM) and explore its effectiveness for producing tunable-fidelity wave functions for use in \textit{ab initio} scattering calculations.  It consists in pre-selecting important configurations based on the value of their contribution to the trace (or, first moment) of the Hamiltonian.  This approach, which we dub configuration-truncation no-core shell model or CT-NCSM, combines the advantages of both the static and dynamic truncation methods: it is interaction-driven as well as flexible in targeting observables other than the energy.

Despite the similarity in name, the NCSM bears little resemblance to the LSSM, and thus it should not be immediately assumed that such an approach will bear fruit. In the LSSM, the low-lying structure of nuclei results from the interactions between configurations of a few valence particles on top of an inert core via an effective residual force, with two-body matrix elements either fitted directly to experimental data~\cite{Brown1988,Honma2002, Brown2006,Nowacki2009,Bouhelal2011,Lubna2020}, or obtained via a renormalization of the nucleon-nucleon (NN) and sometimes three-nucleon (3N) interaction~\cite{HjorthJensen1995,Stroberg2017,Sun2018,Coraggio2024}. 
The NN and 3N interactions that are used in the NCSM typically depend on a significantly smaller number ($\sim$40 compared to the hundreds of matrix elements used in the LSSM) of parameters~\cite{Entem2003,Ekstrom2015,Entem2017,Reinert2018,Jiang2020} and all nucleons are treated as active particles.  

Nevertheless, insights gained from past shell model studies have been found to still be applicable in the NCSM~\cite{Launey2020,Sargsyan2021}. 
Over the years, multiple methods have emerged attempting to tackle the dimensionality explosion problem in the shell model with some even finding application in the NCSM, or vice versa~\cite{Chen1992,Koonin1997,Otsuka2001,Roth2009,Stumpf2016, Launey2020}. 
Most of these truncation methods fall broadly into two categories: methods that statically pre-select the basis on which the many-body wave function will be expanded (as, for example, in Ref.~\cite{Mizusaki2002,McCoy2020,Launey2020,Mercenne2022}), and methods that dynamically select basis states based on the adopted Hamiltonian, such as the IT-NCSM.  
The former class of methods typically still relies on some general aspects of the adopted nuclear interaction and offers the flexibility of selecting basis states that better fit the search for observables other than the energy (e.g., basis of states that are strongly linked to the electric quadrupole operator if a good description of deformation is sought). 
The latter class, being dynamically generated based on a given Hamiltonian, is free from potential biases that may be introduced when pre-selecting the basis 
to target specific observables. 

This paper is organized as follows: In Section II we outline the basics of basis construction in the NCSM, present the systematics of Hamiltonian first moments for various types of nuclear interactions, and briefly revisit the extensions to the NCSM that allow the calculation of nuclear scattering and reaction observables. Then, in Section III we move on to apply the CT-NCSM 
in the calculation of structural properties of light nuclei (specifically $^7$Li) and the scattering of neutrons on $^{12}$C using the nucleon-nucleon interaction described in~\cite{Entem2003}, softened with a similarity renormalization group (SRG) approach. We also present a first demonstration of the reach of the CT-NCSM by computing the differential cross section for the scattering of neutrons from $^{24}$Mg. Concluding remarks and future prospects are given in Section IV.

\section{Methods}
\textit{Ab initio} methods 
solve the quantum mechanical equations for the 
bound state and scattering wave functions arising from the microscopic nuclear Hamiltonian
\begin{align}
    \hat{H} = \hat{T}_{\mathrm{rel}} + \hat{V}^\mathrm{NN}+\hat{V}^\mathrm{3N},
    \label{eq:ham}
\end{align}
with the three operators on the right-hand side denoting the relative kinetic energy between the nucleons, and the NN and 3N components of the nuclear interaction respectively. 
This  Hamiltonian, when projected on a many-body harmonic oscillator basis (as is done in the NCSM), typically leads to binding energies  
that are the result of large cancellations between the expectation values of the kinetic energy and NN interaction, whereas the 3N part of the interaction plays a somewhat more limited role.  

\subsection{No-core shell model}
In the NCSM, the nuclear many-body problem is solved on a complete harmonic oscillator (HO) basis  
defined by the maximum allowed number of quanta of excitation ($N_\mathrm{max}$) from the lowest Pauli-allowed state. Increasing the value of $N_\mathrm{max}$ yields approximately exponentially convergent energies for the states of the system~\cite{Barrett2013,Furnstahl2012}. Therefore, in the traditional NCSM, the variable $N_\mathrm{max}$ solely 
controls the size of the basis for a  
given nucleus, irrespective of the adopted interaction model.  
The unique properties of the HO, allow for both the use of Jacobi relative-coordinate basis states~\cite{Navratil1998,Navratil2000} or single-particle Slater Determinants (SDs) and still preserve the translational invariance of the problem~\cite{Navratil2000-PRL,Navratil2000-PRC}. In this paper we focus on the SD version of the approach, which is computationally more advantageous when exploring the structure and reactions of $p$-shell nuclei.

%\textcolor{red}{\subsection{Further basis truncation}}
%%To overcome this limitation, 
%Various methods for truncating the NCSM basis beyond the $N_\mathrm{max}$ prescription have been introduced \textcolor{red}{over the years}, relying for example on full or approximate symmetries of the nuclear Hamiltonian~\cite{Mizusaki2002,Launey2020,McCoy2020}, or using the Hamiltonian to directly select the relevant components of the basis~\cite{Horoi1994,Roth2009}. Here, we discuss a method that falls under the latter umbrella, but is different than the already established importance truncation NCSM (IT-NCSM)~\cite{Roth2009}. 
%We use the energy centroids (see Eq.~\ref{eq:centroid} in next section) of configurations to pre-select the parts of the NCSM basis that would be most relevant to the description of low-lying states as done in~\cite{Horoi1994} in the case of the traditional shell model. 
%Here, ``low-lying" is to be taken with respect to the full Hamiltonian i.e. millions to billions of states, and could go up to tens of MeVs in excitation energy. In the next section we introduce the formalism and discuss the specifics of the method. 
%
\subsection{Configuration-truncation no-core shell model}
\label{sebsec:CT-NCSM}
The SD basis states are 
constructed from HO single-particle wave functions (or, states) $\phi_i$ bearing quantum numbers $i=\{n\ell jj_zt_z\}$, with $n$ being the number of nodes, $\ell$ the orbital angular momentum, $j$ the total angular momentum, $j_z$ the total angular momentum projection, and $t_z$ the isospin projection respectively. 
The radial part of the single-particle wave functions is further determined by the HO frequency, given in units of energy as $\hbar\omega$. 
Similarly, we define an orbital $\alpha=\{n\ell jt_z\}$ as having the same quantum numbers excluding $j_z$, and thus having a degeneracy of $2j_z+1$. 
One could extend this definition to include the isospin projection quantum number (thus giving each orbital a degeneracy of $2(2j_z+1)$), but this is not done here (i.e., there exist a different set of orbitals for protons and neutrons each). 

Furthermore, we define as a configuration %(or partition) 
an arrangement of nucleons in the various proton and neutron orbitals. Each configuration is characterized by a defined value $N\le N_{\rm max}$ of total HO quanta of excitation and contains multiple SDs, with each SD uniquely belonging to a configuration. 
It is worth noting that SDs do not necessarily have a good total spin ($J$) or isospin ($T$). 
However, since the spin raising and lowering operators do not link different configurations, states with good $J$ quantum numbers can be constructed within a single configuration as linear combinations of SDs belonging only to that configuration. States with good isospin would require the extended definition mentioned above (i.e., excluding the $t_z$ quantum number from the definition of an orbital).

As has been extensively discussed in the  
context of the LSSM~\cite{Senkov2011, Johnson2017}, configurations can be assigned an energy centroid,  
identified as the average trace of the Hamiltonian expanded in the SDs included in the configuration. For example, over a set of $M$ orbitals ($\alpha_i$), each with occupation $n_{\alpha_i}$, we can define a configuration $\kappa=[n_{\alpha_1} n_{\alpha_2} \dots n_{\alpha_M}]$ having dimension $d_\kappa$ with a first moment given by:
\begin{align}
\langle \hat{H} \rangle_\kappa = \frac{1}{d_\kappa}\mathrm{Tr}[\hat{H}]_\kappa.
\label{eq:centroid}
\end{align}
The $\kappa$-configuration trace itself can be computed straightforwardly either by explicitly constructing the Hamiltonian matrix projecting only onto SDs belonging to the configuration, or by simply adding up the diagonal contributions from the various 
components of the Hamiltonian according to~\cite{Senkov2011}:
\begin{align}
    \langle\hat{H}\rangle_\kappa = \frac{1}{4}\displaystyle\sum_{ij} H^\mathrm{2b}_{ij}D^{[ij]}_{\kappa} + 
    \frac{1}{36}\displaystyle\sum_{ijk} H^\mathrm{3b}_{ijk} D^{[ijk]}_\kappa.
\end{align}
Here, the indices $i$, $j$, and $k$ enumerate single particle states and the values $D^{[ijk]}_\kappa$ denote the number of SDs within configuration $\kappa$ that have the states $i$, $j$, and $k$ occupied; the definition of $D^{[ij]}_\kappa$ is analogous. The two-body matrix elements $H^\mathrm{2b}_{ij} = V^{NN}_{ijij}$  
along with their three-body counterparts $H^\mathrm{3b}_{ijk} = V^{3N}_{ijkijk}$ are the only matrix elements 
contributing to the diagonal. Higher moments of the Hamiltonian can also be computed without explicit construction of the matrix~\cite{Wong,Senkov2011}, though their calculation quickly becomes cumbersome.

While the eigenstates of the Hamiltonian will be given by superpositions of the various configurations within a specific $N_\mathrm{max}$ model space, one does not expect that the eigenvector amplitude for such configurations will be uniformly distributed within low-lying states.
Indeed, just by considering the convergence pattern of the energy of low-lying states computed in the NCSM~\cite{Barrett2013}, it is clear that configurations with lower values of $N$  
are more important  
as the energy gain from  
increasing  
$N_\mathrm{max}$ values quickly diminishes. As a result, it is not uncommon for the contribution of the $N_\mathrm{max}=0$ part of the wave function to be $\sim 70\%$ in low-lying states. Along this line of reasoning, the centroid energy of each configuration provides a first-order estimate of the energy at which that configuration is expected to become important.  
However,  
the correlation energy contributed by the interaction between configurations will shift states towards lower energies.

\begin{figure}[t]
\centering
\includegraphics[width=0.48\textwidth]{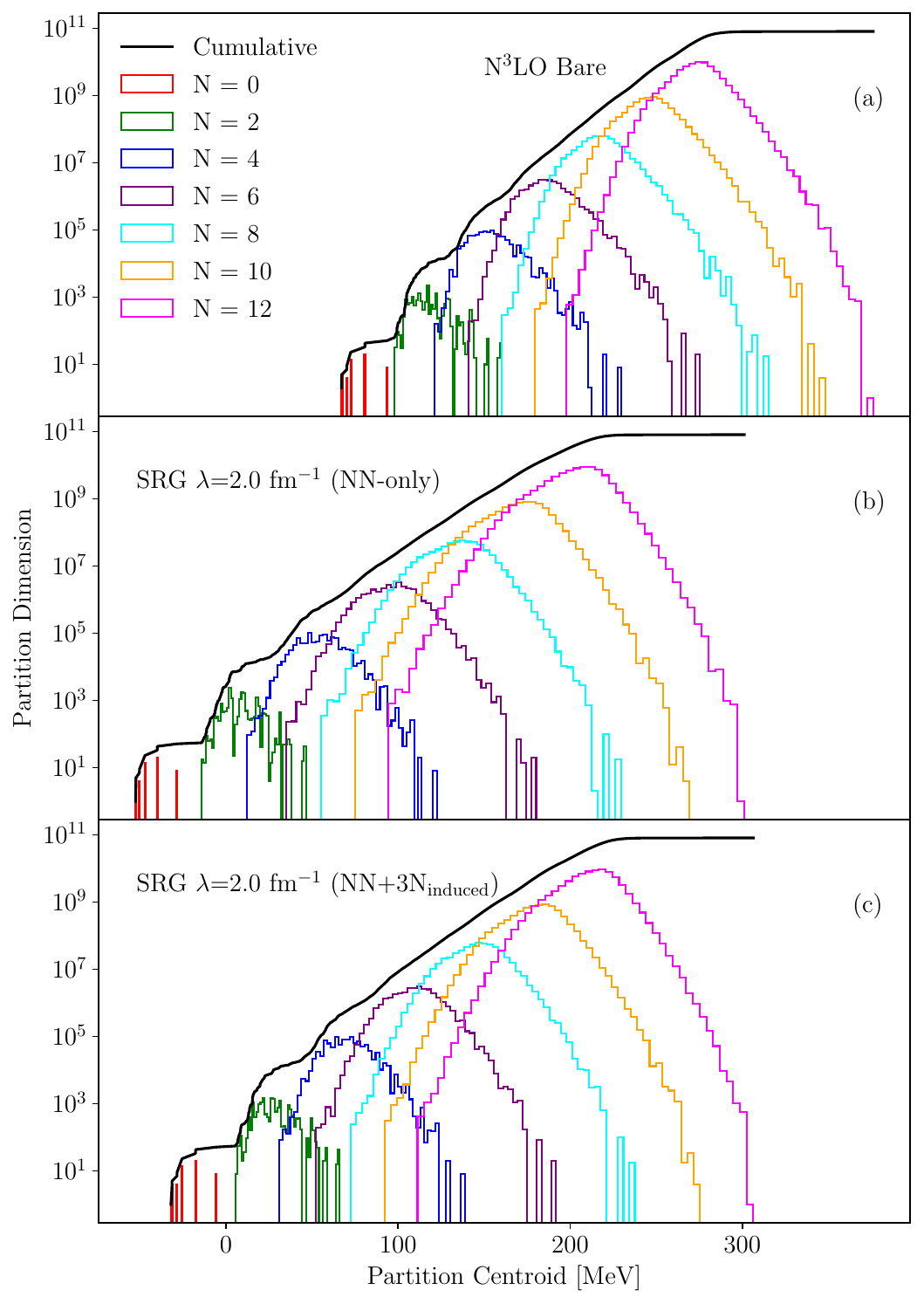}
\caption{Configuration distributions for $^{12}$C. Each group of configurations having the same number of total quanta $(N)$ is plotted against its centroid energy (histograms, each including a few configurations in each bin). The cumulative number of states increases approximately exponentially after $N$=4 until it reaches a plateau since the $N$=14 configurations are not calculated. By setting a target total number of single-particle states the number of included configurations can be truncated without affecting low-lying states.}
\label{fig:centroids12C}
\end{figure}

In the CT-NCSM we thus follow the prescription suggested in Ref.~\cite{Horoi1994} and truncate the basis by selecting a desired total basis dimension and including only the configurations with the lowest centroid energies up to the point where the desired limit (denoted by $N_\mathrm{SD}$) is reached. Since the number of SDs that belong to a configuration is significantly smaller than the full size of the basis, this approach transforms the basis dimension into a quasi-continuous variable, with the (severely reduced in dimension) matrix diagonalization resulting in a tunable-fidelity wave function for the low-lying nuclear states.

Previous work~\cite{Johnson2017} exploited 
this configuration decomposition in the NCSM to 
examine the effects of the SRG evolution of the NN interaction in two-body space (or NN-only). It was found that the SRG transformation results in a lowering of the energy centroids for all configurations accompanied by a spreading of the centroids within sub-spaces with fixed total number of excitation quanta $N$. Repeating  
this study, we can reproduce the behavior and further demonstrate how the inclusion of the  
3N force induced by the SRG evolution of the NN interaction in three-body space, or NN+3N$_{\rm ind}$, somewhat 
alleviates, but does not completely correct, the  
spreading effect 
%resulting in what amounts to an almost uniform shift in centroid energy 
(see Fig.~\ref{fig:centroids12C} and Table~\ref{tab:cents}). 

\begin{figure}[t]
\centering
\includegraphics[width=0.47\textwidth]{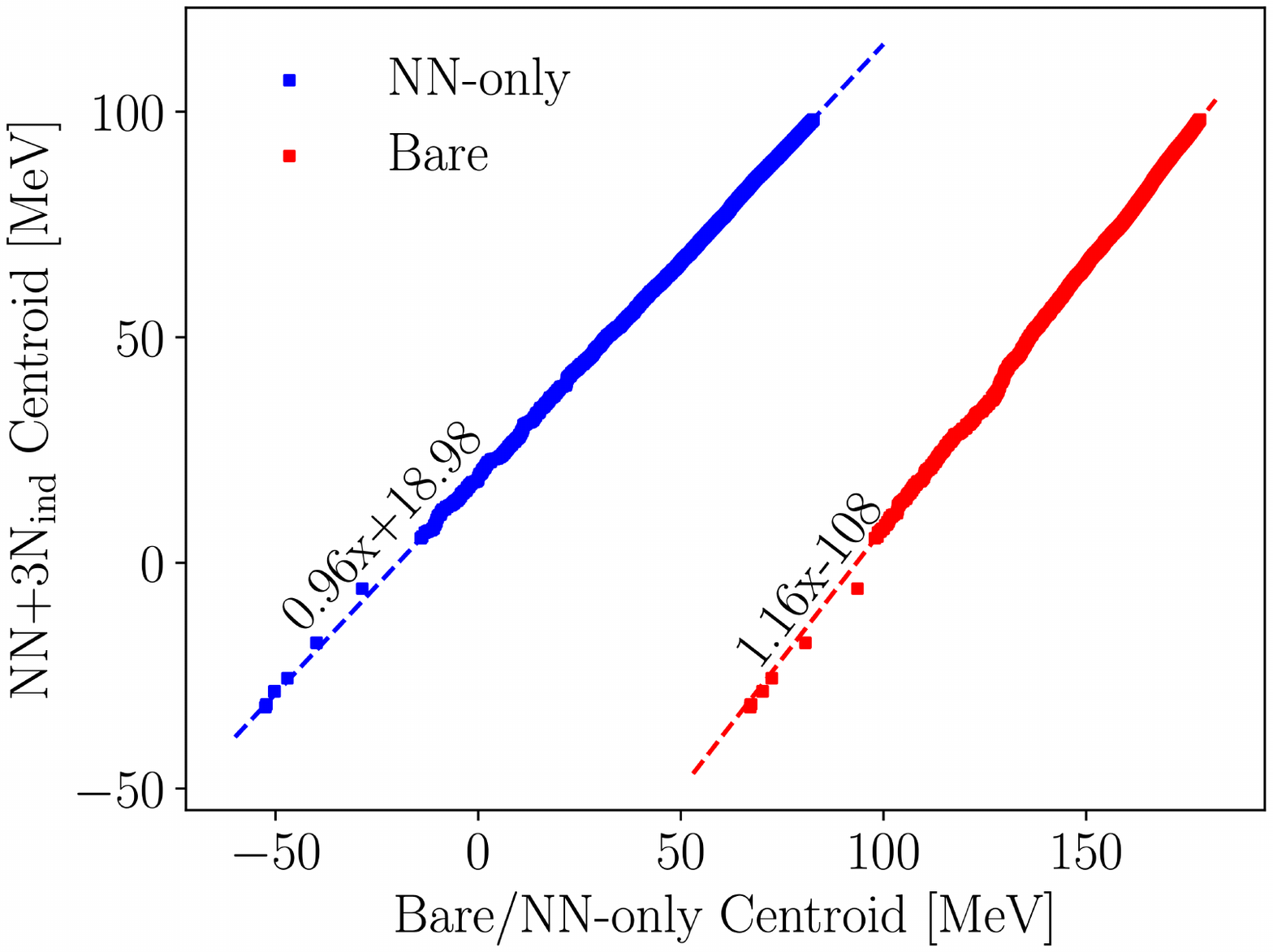} %Note this is png because pdf takes too long to render
\caption{Centroid energies of the first 2$\times$10$^4$ configurations using the NN-only and Bare interactions plotted against those obtained with the NN+3N$_\mathrm{ind}$ interaction for $^{12}$C. The absence of any obvious zig-zag shape in the data points to the lack of sensitivity of configuration ordering to the specifics of the interaction. Linear fits are shown to guide the eye.}
\label{fig:CentroidChoicePlot}
\end{figure}

Furthermore, Fig.~\ref{fig:centroids12C} 
clearly shows that the specifics of the interaction are not essential in making a selection of the most important configurations.  
Indeed, 
for a large enough value of $N_\mathrm{SD}$, the last few configurations to be included should not play an important role in the  
description of the wave function. 
In addition,  
the ordering of configurations shown in each of the three panels of Fig.~\ref{fig:centroids12C}
hardly differ macroscopically (i.e., apart from small re-orderings among a few configurations). 
This observation is also supported by the absence of zig-zagging in the energy centroids obtained with the NN+3N$_{\rm ind}$ interaction plotted versus those from the NN-only and bare interactions
(see Fig.~\ref{fig:CentroidChoicePlot}). 
%It is worth pointing out that the slope of the fit to the NN-only calculation is closer to unity than that of the unevolved (bare) interaction. Naively, one could expect that the inclusion of the induced force recovers exactly the same properties of the bare interaction, i.e. the configurations are simply shifted by a constant. This does not appear to be the case, with the low-lying configurations all being below the fit line, indicating that under SRG evolution with the given generator (the relative kinetic energy) it is the lowest ($N=0$) configurations that shift further than the rest.

\begin{table}[t]
    \centering
    \begin{tabular}{c c c c}
        Quanta (N)&	Bare&	NN-only&	NN+3N$_\mathrm{ind}$\\
        \hline\hline
0&	(78.3, 8.11)&	(-42.2, 7.3)&	(-20.2, 7.8)\\
2&	(118.2, 10.1)&	(9.1, 9.7)&	(27.8, 9.7)\\
4&	(153.8, 11.2)&	(54.8, 12.1)&	(70.6, 11.4)\\
6&	(186.7, 11.8)&	(96.6, 13.6)&	(109.8, 12.6)\\
8&	(217.4, 12.0)&	(135.0, 14.5)&	(146.2, 13.3)\\
10&	(246.4, 11.9)&	(170.7, 14.8)&	(180.2, 13.6)\\
12&	(273.9, 11.7)&	(204.1, 14.9)&	(212.2, 13.6)\\
\hline
    \end{tabular}
    \caption{Means and variances ($\mu,\sigma$) for each set of configurations with a fixed total number of quanta (N) for each of the three interactions shown in Figure~\ref{fig:centroids12C}.}
    \label{tab:cents}
\end{table}

We find a similar ordering with the inclusion of various forms of three-nucleon forces~\cite{Navratil2007,Gazit2019,Soma2020}, again pointing to the lack of any significant reorganization arising from the specifics of the 3N part of the interaction.

\subsection{No-core shell model extensions for \textit{ab initio} calculations of scattering and reactions} 
\label{subsec:ncsmcont}

The NCSM has been extended to the description of scattering through a variety of methods~\cite{Quaglioni2008,Baroni2013L,Papadimitriou2013,Kravvaris2017,Shirokov2022}. Most relevant to this work, the NCSM combined with the resonating group method (NCSM/RGM)~\cite{Quaglioni2008} and the NCSM with continuum (NCSMC)~\cite{Baroni2013L} rely on the ability to explicitly construct microscopic binary-cluster states where the two reactants (arranged in a total angular momentum-parity $J^\pi$ channel and places a distance $r$ from each other)  are described by NCSM wave functions:
\begin{align}
\label{eq:inteq}
	% \left| \Psi^{J^\pi}\right\rangle = \sum_{\nu}\int dr r^2 \frac{\gamma_{\nu}^{J^\pi}(r)}{r} \left|\Phi^{J^\pi}_{\nu r}\right\rangle,
 \nonumber 
 \left|\Phi^{J^\pi}_{\nu r}\right\rangle = 
&\left[
    \left(
        \left|
            A-a\alpha_tJ_t^{\pi_t}
        \right\rangle
        \left|
            a\alpha_pJ_p^{\pi_p}
        \right\rangle
    \right)^{(s)}
 \right.\\
&\left.\times Y_\ell(\hat{r}_{A-a,a})
 \right]^{J^\pi} \frac{\delta(r-r_{A-a,a})}{rr_{A-a,a}}.
\end{align}
The index $\nu$ contains all other quantum numbers needed to define the the basis states, namely the masses $A-a$ and $a$ and spin-parities  of the target ($I^{\pi_t}_t$) and projectile ($I^{\pi_p}_p$) nucleus, respectively, (any other quantum number needed to define the target/projectile internal states is given in the collective index $\alpha_{t/p}$)as well as asymptotic quantum numbers $\ell$ and $s$ determining, respectively, the relative angular momentum and spin of the reaction channel. The total wave function of the system is then determined as an expansion over all the $\left|\Phi^{J^\pi}_{\nu r}\right\rangle$ states
\begin{align}
    \left|\Psi_\mathrm{NCSM/RGM}^{J^\pi}\right\rangle = \displaystyle\sum_\nu\int dr r^2 \frac{\gamma_\nu^{J^\pi}(r)}{r} \left|\Phi^{J^\pi}_{\nu r}\right\rangle,
\end{align}
with the continuous amplitudes $\gamma_\nu^{J^\pi}(r)$ determined using the microscopic R-matrix method on a Lagrange mesh~\cite{Hesse1998,Descouvemont2010}.

The NCSMC further extends this description by augmenting the binary-cluster basis with NCSM-computed states of the aggregate system (here, with mass $A$) 
\begin{align}
    \left|\Psi_\mathrm{NCSMC}^{J^\pi}\right\rangle = \displaystyle\sum_\lambda c_\lambda\left|A\lambda J^\pi\right\rangle + \displaystyle\sum_\nu\int dr r^2 \frac{\gamma_\nu^{J^\pi}(r)}{r} \left|\Phi^{J^\pi}_{\nu r}\right\rangle ,
\end{align}
where $\lambda$ enumerates the various $A$-body eigenstates obtained in the NCSM, and the coefficients $c_\lambda$ are determined simultaneously with the $\gamma^{J^\pi}_\nu(r)$ amplitudes.

The computational difficulty in both of these methods arises in computing the matrix elements of the many-body Hamiltonian between the binary-cluster states, a difficulty further exacerbated with increasing $N_\mathrm{max}$ values for the wave functions of the fragments and, as a consequence, increasing the number of SDs in the expansion. Naturally, combining the CT-NCSM with both the NCSM/RGM and the NCSMC is expected to significantly alleviate this issue, and could open a path to the \textit{ab initio} description of reactions for nuclei beyond oxygen. 

\section{Results}
To demonstrate the performance of the CT-NCSM, we conducted NCSM, NCSM/RGM and NCSMC calculations starting from a microscopic two-nucleon Hamiltonian. That is, for the present study, we disregard the three-nucleon force term of Eq.~(\ref{eq:ham}), as well as the SRG-induced 3N components. There is no issue in generalizing the method to 3N forces.
Specifically, we adopt the interaction described in~\cite{Entem2003}, evolved to an SRG resolution scale of $\lambda=2$~fm$^{-1}$~\cite{Jurgenson2013}, and set the oscillator parameter for the NCSM basis at $\hbar\omega$=20~MeV as these are typical choices used in many light-nuclei calculations in the past~\cite{Hupin2015,Hebborn2022,Kravvaris2023}. 

\subsection{Low-lying spectrum of $^7$Li}
We start by considering the nucleus of $^7$Li. This nucleus is light enough that NCSM calculations can be performed up to $N_\mathrm{max}$=12 with relative ease, but complex enough that its  spectrum of low-energy levels is affected by multiple modes of clusterization ($^3$H+$^4$He, $n$+$^6$Li, and eventually $p$+$^6$He). 
Reproducing the co-existence of such modes in the CT-NCSM is a necessary component in demonstrating the effectiveness of the present approach for the description of any nuclear state, including those that may present different properties (or share  different dominant configurations) from the ground state.
% For example, if one where to construct a rotational band built on the $3/2^-$ ground state, the pair of $5/2^-$ states at $\sim7$ MeV of excitation energy would not be properly reproduced as one of them has a relatively small B(E2) value for the transition to the ground state.

\begin{figure}[!t]
\centering
\includegraphics[width=0.47\textwidth]{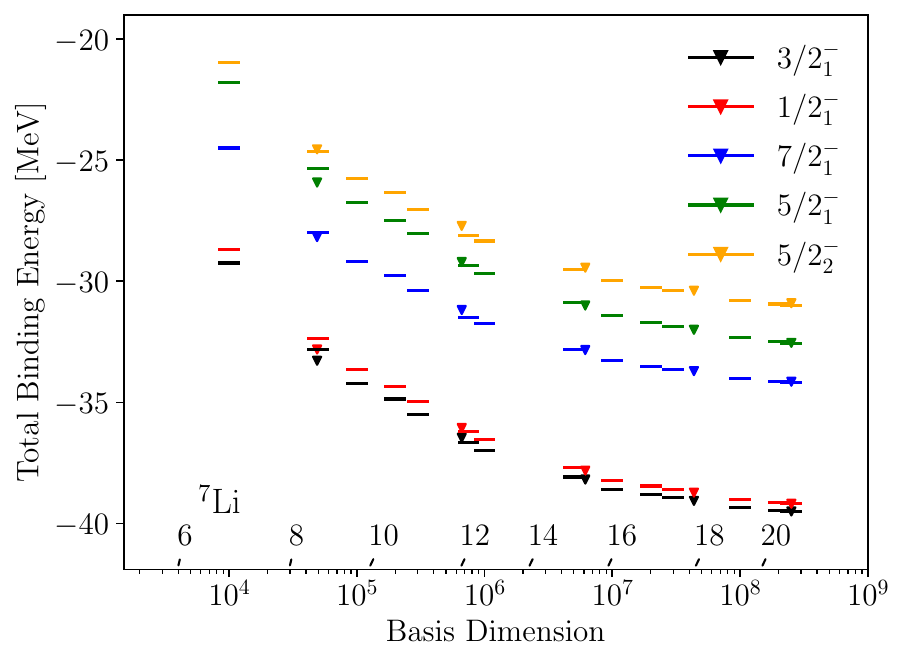}
\caption{Convergence of absolute energies of low-lying states in $^7$Li with respect to the CT-NCSM basis size, shown by horizontal bars. Triangles correspond (going from the left to the right in increasing size of the basis) to energies obtained with the traditional NCSM $N_\mathrm{max}=4$, 6, 8, 10, and 12  truncation. The markings along the horizontal axis indicate at which basis dimension the first configuration corresponding to a specific total number of quanta enters. }
\label{fig:Li7cvg}
\end{figure}

First, we look at the convergence of the $^7$Li spectrum of energy levels with respect to the size of the CT-NCSM basis ($N_\mathrm{SD}$). 
As described in the previous section, we use the first moment of the Hamiltonian as a guide to order the configurations. We progressively truncate the many-body basis at increasing values of $N_\mathrm{SD}$, ranging from $\sim$10$^4$ SDs to $\sim$10$^8$ SDs, by correspondingly retaining an increasing number of configurations. 
We consider here all configurations having a total number of quanta $N\leq 20$; for reference, the full $N_\mathrm{max}$=20 space would contain roughly 7.5$\times$10$^{10}$ SDs which, to the best of our knowledge, is a factor of 3 larger than currently accessible dimensions~\cite{Forssen2018}. 
As a reminder, in the $N_\mathrm{max}$ truncation typically adopted in the NCSM, all and only states with up to a fixed total number of quanta are included; in the present configuration truncation scheme, there is no such restriction with, for example, basis states reaching as high as $N_\mathrm{max}$=18 being included already at $N_\mathrm{SD}$=4.3$\times$10$^7$ (see  Fig.~\ref{fig:Li7cvg}), a dimension similar to that of the full $N_\mathrm{max}$=10 space.

The convergence of the CT-NCSM absolute energies in the low-lying spectrum of $^7$Li follows a similar exponential convergence as observed in the LSSM and the NCSM~\cite{Horoi1994, Furnstahl2012} with increasing $N_\mathrm{SD}$ (Fig.~\ref{fig:Li7cvg}). 
Moreover, the NCSM calculations obtained with the  $N_\mathrm{max}$ truncation mostly lie along the same convergence pattern as those obtained with the CT-NCSM despite of the different HO quanta content of the basis states included.
%: in the $N_\mathrm{max}$ truncation, all and only states with up to a fixed total number of quanta are included; in the present configuration truncation scheme, there is no such restriction with, for example, basis states reaching as high as $N_\mathrm{max}$=18 being included already at dimension 43$\times10^6$ (see  Fig.~\ref{fig:Li7cvg}).

%
%
%
That we do not find a substantially accelerated convergence is not unexpected; the fact that in many-body calculations a large part of the total binding energy comes from many states with small amplitudes 
has been demonstrated in various cases over the years~\cite{Kramer,Kravvaris2019}. 
While in past microscopic approaches some part of the interaction could be tuned to reproduce the total binding energy for a given wave function truncation scheme~\cite{Damman2009,Dubovichenko2023}, this is not an easily accomplished--or desired--feat in \textit{ab initio} calculations. 
It is, however, helpful to have a truncation scheme that follows exponential convergence, as absolute energies can be obtained by extrapolation. In this respect, the CT-NCSM is highly favorable as the energy of a given state can be computed at additional basis dimensions compared to the traditional $N_\mathrm{max}$ truncation (see. e.g., Fig.~\ref{fig:Li7cvg}), thus providing an increased data set for constraining such an extrapolation.

The wave functions obtained with such successively larger bases provide further insight on the convergence of the CT-NCSM. 
At roughly $N_\mathrm{SD}$=3$\times$10$^7$ we obtain a $\sim$1\% change in binding energy compared to $N_\mathrm{SD}=10^7$, hitting the point of diminishing returns. 
For all five states in Fig.~\ref{fig:Li7cvg}, the difference between the energies obtained with $N_\mathrm{SD}=10^8$ versus $N_\mathrm{SD}=10^7$ is $\sim 2\%$ and the overlap between the corresponding wave functions is greater than 98.6\%, corroborating that reasonable convergence has been achieved.  %meaning their bulk character should be reasonably converged.  
Remarkably, the $5/2_2^-$ state is also well-reproduced in the CT-NCSM, despite it belonging to a sufficiently different class of states than the ground-state rotational band. This serves as a demonstration that the truncation scheme underlying the CT-NCSM is unbiased and effective. 

Next, we investigate the convergence properties of the matrix elements of electromagnetic transition operators computed within the low-lying $^7$Li states obtained within the CT-NCSM.
Naively, one may think that the inclusion of basis states with higher quanta earlier on in the many-body basis would provide an improved description of the long-range part of the wave functions and thus an accelerated convergence of  the electromagnetic transitions computed from them, such as B(E2) quadrupole transition probabilities. 
However, (relatively large) collective transition rates 
emerge from multiple coherent transitions over all the SDs 
with small amplitudes and thus B(E2) values computed within the CT-NCSM follow the same convergence pattern as they would using the standard $N_\mathrm{max}$ truncation (see Table~\ref{tab:BE2rates}). 

\begin{table}[t]
\centering
\setlength{\tabcolsep}{8pt}
\renewcommand{\arraystretch}{1.5}
\begin{tabular}{r l l l l}
Dimension  & \multicolumn{4}{c}{Initial State}\\
$(\times10^6)$ & 1/2$^-_1$  & 7/2$^-_1$ & 5/2$^-_1$ & 5/2$^-_2$ \\[0.5em]
\hline\hline
0.66$^*$                     & 5.349 & 2.550 & 0.654 & 0.010 \\
0.75\textcolor{white}{$^x$}  & 5.139 & 2.464 & 0.621 & 0.013 \\
5\textcolor{white}{$^x$}     & 5.891 & 2.879 & 0.730 & 0.011 \\
6.2$^{*}$                    & 6.256 & 3.069 & 0.790 & 0.009 \\
30\textcolor{white}{$^x$}    & 6.728 & 3.352 & 0.860 & 0.011 \\
43.6$^*$                     & 7.087 & 3.542 & 0.921 & 0.010 \\
100\textcolor{white}{$^x$}   & 7.360 & 3.714 & 0.961 & 0.010 \\
\hline
\end{tabular}
\caption{B(E2) values (in $e^2$fm$^4$) for excited states of $^7$Li decaying into the 3/2$^-$ ground state. Asterisks indicate spaces where the SDs were selected using the standard $N_\mathrm{max}$ truncation (here, in descending order, $N_\mathrm{max}=6,8,10)$. }
\label{tab:BE2rates}
\end{table}

Finally, 
one of the advantages of using the $N_\mathrm{max}$ truncation is that it enables the exact separation of the center-of-mass (CM) part of the wave function~\cite{Barrett2013}. \textit{A priori}, the CT-NCSM 
does not enjoy this property because the CM Hamiltonian ($H_\mathrm{CM}$, here defined so as to be dimensionless) mixes different configurations. In principle, one could  
extend the configuration truncation scheme to include all configurations required to exactly diagonalize $H_\mathrm{CM}$. Since $H_\mathrm{CM}$ is a two-body operator, if a configuration is selected then all configurations that change the orbital occupations by up to a number of two particles would need to also be included. However, such an approach would severely limit the benefits of the CT-NCSM, quickly leading to large basis dimensions. Rather, we use a technique almost as old as the shell model itself~\cite{Palumbo1967,Palumbo1968,Gloeckner1974} that consists in diagonalizing the altered Hamiltonian 
\begin{align}
    H^\beta = H + \beta\times \left(H_\mathrm{CM} -\tfrac{3}{2}%\hbar\omega
    \right),
    \label{eq:HwithNotLawsonTerm}
\end{align}
where $\beta$ is chosen to be some large value (we choose here a value of 40 MeV). 
The effect of adding this term is to separate (spurious) eigenstates that contain non-zero-quanta CM contributions and shift them to higher energies.
We find that the number of CM quanta in our final wave functions never grow above 0.002 for calculations with a reasonably large basis size. (Very small values for $N_\mathrm{SD}$ tend to yield up to 0.005 CM quanta.) 
%\textcolor{red}{Will double-check this.} -- Checked and it holds up, 10k basis for 7Li has 0.0048 CM quanta. 
Note that these values concern wave functions obtained using the hamiltonian of Eq.~(\ref{eq:HwithNotLawsonTerm}) and thus provide a good measure of the spuriosity contributions as discussed in Ref.~\cite{Roth2009cm}. 
One could be tempted to use the altered 
Hamiltonian of Eq.~\eqref{eq:HwithNotLawsonTerm} already when selecting the configurations, however this would simply result in a spreading (dependent on the value of $\beta$) of configurations having different total numbers of quanta~\cite{Horoi1994} because the selection process utilizes only the diagonal part (trace) of the Hamiltonian. 

\subsection{Low-energy neutron-$^{12}$C scattering}
Keeping in mind that the main driver for this work, we now turn to the scattering of a neutron from $^{12}$C. 
The two methods that will be employed are the NCSM/RGM~\cite{Quaglioni2008,Quaglioni2009,Navratil2016} and the NCSMC~\cite{Baroni2013L,Baroni2013C,Navratil2016}, briefly introduced in Section~\ref{subsec:ncsmcont}. 
While the NCSM/RGM requires a good description of the target nucleus $^{12}$C, the NCSMC further requires as input NCSM wave functions of the aggreagate projectile-target system (in this case $^{13}$C). 
In both methods we replace all NCSM-obtained wave functions (i.e., for states of $^{12,13}$C) with CT-NCSM ones and study the convergence with respect to $N_\mathrm{SD}$. 
The configuration centroid structure for $^{12}$C (Figs.~\ref{fig:centroids12C} and ~\ref{fig:CentroidChoicePlot}) has already been discussed in Sec.~\ref{sebsec:CT-NCSM} 
and we do not repeat it here; the ground state energies obtained for both nuclei at various values of $N_\mathrm{SD}$ are listed in Table~\ref{tab:carbonconvergence}. 
Note that calculations of different nuclei with the same $N_\mathrm{SD}$ should not be directly compared to evaluate, for example, the position of decay thresholds, as the fidelity obtained for each nucleus is different. 
In traditional NCSM calculations, this comparison is done on the basis of $N_\mathrm{max}$; while \textit{a priori} there is no particular reason for which energies obtained for different nuclei at the same $N_\mathrm{max}$ value should yield the best description of reaction thresholds, in practice such a behavior has been observed~\cite{Hupin2015}.

\begin{table}
    \centering
    \begin{tabular}{c c c c c c}
        $N_\mathrm{SD}$&	\multicolumn{3}{c}{CT-NCSM}  &	\multicolumn{2}{c}{CT-NCSM/RGM} \\
        & $^{12}$C$_\mathrm{g.s.}$ & $^{13}$C$_{1/2^-}$ & $^{13}$C$_{1/2^+}$ & $^{13}$C$_{1/2^-}$ & $^{13}$C$_{1/2^+}$\Bstrut\\
        \hline\hline
3$\cdot 10^7$  &	-96.38&	-104.73 & -93.26 & -3.15 & -0.23\Tstrut\\ %& -99.53  & -96.61\\ %
5$\cdot 10^7$  &	-97.08&	-105.58 & -94.40 & -2.98 & -0.22\\ %& -100.06 & -97.30\\ %
7.5$\cdot 10^7$&	-97.57&	-106.01 & -95.19 & -2.72 & -0.22\\ %& -100.29 & -97.78\\ %
1$\cdot10^8$   &	-97.87&	-106.32 & -95.84 & -2.63 & -0.22\\ %& -100.50 & -98.09\\ %
1.5$\cdot10^8$ &	-98.36&	-106.72 & -96.69 & -2.54 & -0.22\\ %& -100.90 & -98.58\\ %
2$\cdot10^8$   &    -98.64&	-107.01 & -97.25 & -2.49 & -0.22\\ %& -101.13 & -98.86\\ %
2.5$\cdot10^8$ &    -98.86& -107.24 & -97.68 & -2.43 & -0.22\Bstrut\\ %& -101.28 & -99.07\\ %
\hline
Extrap.        &    -98.93& -107.32 & -97.83 &         &      \Tstrut\\
Exp.      &   -92.16&  -97.11  & -94.02  \\
\hline\hline
    \end{tabular}
    \caption{Absolute values for the energies of $^{12,13}$C in the CT-NCSM and $^{13}$C binding energies w.r.t. the $n$+$^{12}$C threshold in the CT-NCSM/RGM for various basis sizes ($N_\mathrm{SD}$). Note that the overbinding compared to experiment is a byporduct of the NN-only SRG evolved interaction used in the calculations, as discussed for example in Ref.~\cite{Navratil2010}. }
    \label{tab:carbonconvergence}
\end{table}
Apart from the $N_\mathrm{max}$ (or $N_\mathrm{SD}$) used to describe the wave function of the target (and projectile if it is a composite particle) an additional parameter that defines (CT-)NCSM/RGM and (CT-)NCSMC calculations is the size of the harmonic oscillator model space ($N_\mathrm{rel}$) used to represent the relative motion  when computing the (projectile-target)  potential kernel~\cite{Quaglioni2008,Quaglioni2009,Kravvaris2020a}. 
%Since in the NCSM/RGM the potential is expanded in an harmonic oscillator basis, the maximum number of quanta allowed in this expansion ($N_\mathrm{rel}$) corresponds to this model parameter. 
In the traditional approach, $N_\mathrm{rel}$ is set equal to the maximum shell that particles in the target nucleus can reach for a given $N_\mathrm{max}$ value, incremented by one to account for the parity opposite to that of the ground state (so for $s$-shell nuclei $N_\mathrm{rel}=N_\mathrm{max}+1$, and for $p$-shell nuclei $N_\mathrm{rel}=N_\mathrm{max}+2$). 
Using this approach may lead to ambiguities when comparing standard NCSM/RGM calculations with those performed with the bases selected using the CT-NCSM, since states with different total numbers of quanta would appear at different bases sizes (see Fig.~\ref{fig:Li7cvg}). 
%The result then could be abrupt changes in an otherwise smooth convergence process as the basis size is increased once the introduction of a single configuration that contains a larger number of total quanta is reached. 
Rather, in the following we set $N_\mathrm{rel}=14$ for all calculations, including those in which the $^{12}$C wave function is obtained with the traditional $N_\mathrm{max}$ truncation scheme. 
We choose this value because the largest CT-NCSM basis used in the present study was constructed using states coming from configurations with up to $N_\mathrm{max}=12$ excitation quanta. 

\begin{figure}[t]
\centering
\includegraphics[width=0.47\textwidth]{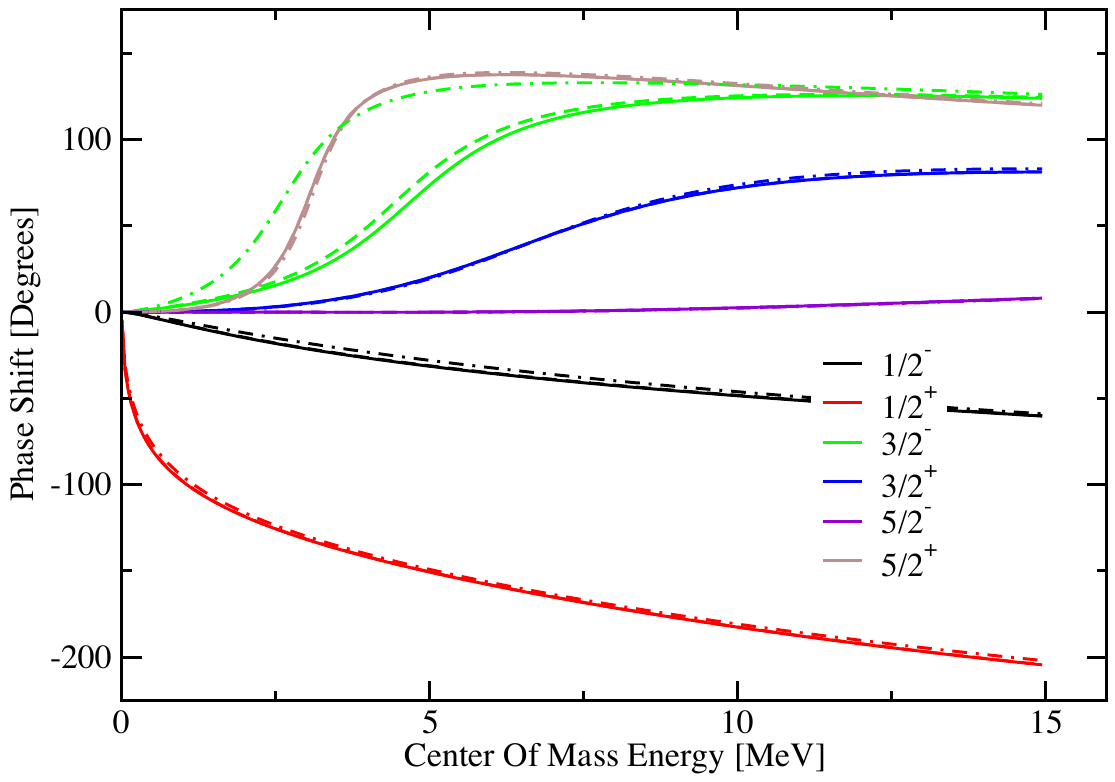} 
\caption{$^{12}$C+$n$ phase shift convergence with increasing size of the selected basis. Dot-dashed, dashed, and solid lines correspond to calculations using $N_\mathrm{SD}=10^6$, 3$\times 10^7$, and 5$\times 10^7$ many-body states respectively. The $^{12}$C+$n$ relative motion is expanded on an $N_\mathrm{rel}=14$ space in all cases.}
\label{fig:ncsmrgm_12C}
\end{figure}

The neutron scattering phase shifts computed within the CT-NCSM/RGM show rapid convergence even with a very modest number of basis states (Fig.~\ref{fig:ncsmrgm_12C}). 
In all partial waves, lest for the 3/2$^-$ case, there is virtually no difference by almost doubling the size of the basis from $N_\mathrm{SD}$=3$\times$10$^7$ to 5$\times$10$^7$, with a small but somewhat visible difference when comparing to calculations using $N_\mathrm{SD}=10^6$. 
It is worth noting the remarkable convergence of all phase shift channels other than the $3/2^-$ which should clearly be attributed to fixing $N_\mathrm{rel}$ at a relatively large value. 
In these cases it would appear that the specifics of the target state are not important overall, however, the RGM norm and Hamiltonian kernel matrix elements~\cite{Quaglioni2009} vary at the order of $\sim 1-10\%$ between the calculation with $N_\mathrm{SD}$=10$^6$ and the one with $N_\mathrm{SD}$=5$\times 10^7$. 

\begin{figure}[t]
\centering
\includegraphics[width=0.47\textwidth]{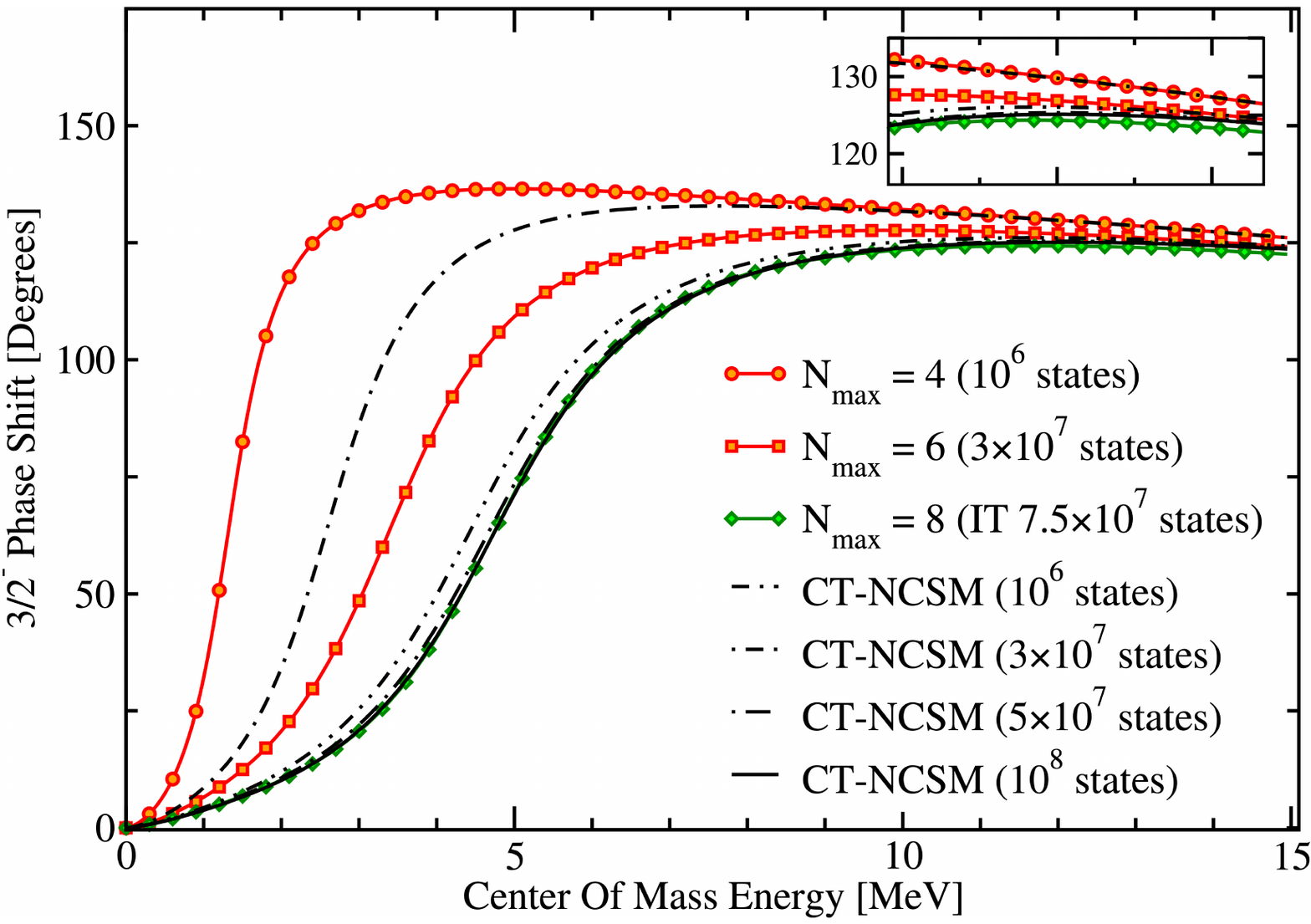} 
\caption{$^{12}$C+$n$ phase shifts in the 3/2$^-$ channel obtained with the CT-NCSM scheme (black lines) compared to the ones obtained with the $N_\mathrm{max}$ scheme (red lines with points) of approximately the same dimension, as well as with wave functions obtained in the $N_\mathrm{max}$=8 IT-NCSM with $\kappa=10^{-5}$ (see Ref~\cite{Navratil2010} for details). The number in parentheses for each space indicates their respective dimension. We observe a rapid convergence using the CT-NCSM. The inset shows the higher-energy phase shifts, highlighting the discrepancy between the CT-NCSM and IT-NCSM results, likely originating in the $N_\mathrm{max}>8$ parts of the $^{12}$C wave function probed in the CT-NCSM. All calculations where done in the same $N_\mathrm{rel}=14$ space for the carbon-neutron relative motion.}
\label{fig:NmaxTruncationComparison}
\end{figure}

Within the CT-NCSM/RGM, we also compute the $^{13}$C ground (1/2$^-$) and first excited (1/2$^+$) bound states; no other bound states are found. 
The ground state is found to be bound by about 2.5 MeV for the largest numbers of basis states considered in the CT-NCSM calculation of the $^{12}$C ground state (see Table~\ref{tab:carbonconvergence}). 
The first excited state binding energy is remarkably better converged with respect to the $^{12}$C+$n$ threshold, having converged to a value of 0.2 MeV even in calculations with smaller bases ($N_\mathrm{SD}<10^7$). 
It should be noted that both these binding energies come as differences between the energy of the $^{12}$C wave function used to describe the target and the ones obtained from the CT-NCSM/RGM calculation for $^{13}$C. Thus they represent a fairly delicate cancellation between large numbers. 
The different convergence patterns for these two states, however, demonstrates how extending the many-body basis with the microscopic binary-cluster states of Eq.~(\ref{eq:inteq}) affects nuclear properties at different scales.

On the one hand, the 1/2$^+$ state--that evolves to be the ground state in heavier carbon isotopes--appears at a lower energy than the 3/2$^-$ state as in experiment, showing its halo-like structure is indeed captured in the truncated calculations as was the case for the second 5/2$^-$ excited state in $^7$Li. 
As a consequence of this structure, the state converges almost immediately once the binary-cluster channels are included in the wave function description.
On the other hand, the ground state of $^{13}$C is still a compact state with its wave function spread over multiple SDs, and the short-range correlations that are further incorporated with increasing $N_\mathrm{SD}$ are more important.

%As seen by the binding energies in the previous paragraph, within the NCSM/RGM, the absolute binding energy of the ground state shows a similar convergence pattern as in an NCSM calculation, gaining roughly 3 MeV of binding compared to the NCSM calculation for $^{12}$C regardless of basis size. 
%The convergence properties discussed here lead us to exactly the same conclusions as before when it comes to the effects of the ``grass," with little difference is found in the RGM Hamiltonian kernel. \textcolor{red}{Revisit. Perhaps a plot with ANCs vs binding energy w.r.t. threshold would also be instructive (annotations to show basis dimension for each point).}

The rapid convergence of the 3/2$^-$, 3/2$^+$, and 5/2$^+$ resonances further indicates that the structure of these states should also be dominated by the binary-cluster channel components. 
Like the 1/2$^+$ state, the 3/2$^-$ and 5/2$^+$ states correspond to experimentally bound levels, but to accurately describe their position with respect to the $n$+$^{12}$C threshold within the CT-NCSM/RGM it is possible that calculations including the $2^+$ excited state of the $^{12}$C target would need to be performed, preferably including the SRG-induced 3N part of the NN interaction.
%In past work~\cite{Baroni2013L,Baroni2013C}, it was shown that the inclusion of NCSM-like states of the aggregate nucleus (here $^{13}$C) leads to an improved description of short-range properties of the wave function, as well as the position of bound states and resonances. 
% Since these aggregate states are also built within the NCSM, it is also straightforward to apply the CT-NCSM to them as well. 

The comparison with calculations using the $N_\mathrm{max}$ truncation scheme is also quite favorable. 
We find accelerated convergence with the same total size of basis indicating that the configuration centroid selection scheme succeeds in capturing the essential features of the target wave function with a reduced basis size (see Fig.~\ref{fig:NmaxTruncationComparison} for a comparison of the least-well-converged 3/2$^-$ phase shifts). 
Furthermore, the good agreement between calculations using the $N_\mathrm{max}$ truncation and the one presented here point to only minimal effects from the small CM contamination of the target wave function.

\begin{figure}[t]
\centering
\includegraphics[width=0.47\textwidth]{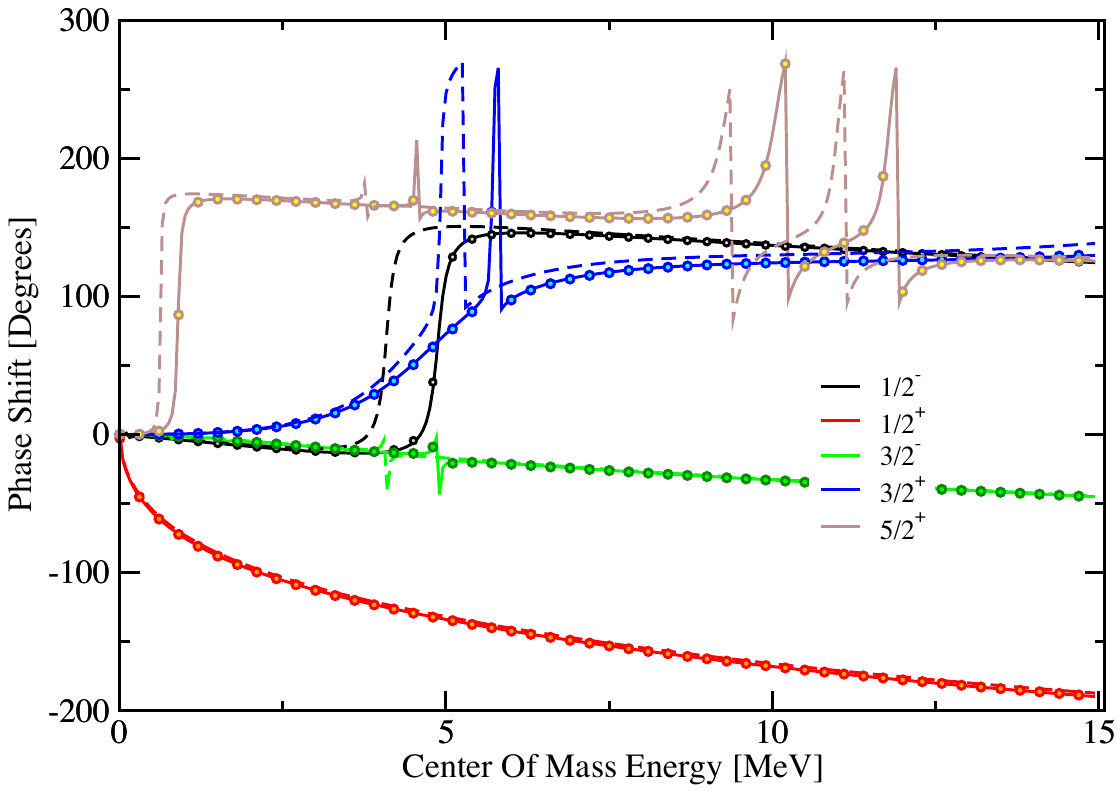} 
\caption{NCSMC-computed $^{12}$C+$n$ phase shifts using different fidelity wave functions--5$\times$10$^7$ (dashed) and $10^8$(solid)--for the NCSM/RGM components and $10^8$ states for both negative and positive parity states of $^{13}$C. 
Points indicate phase shifts for the 5$\times$10$^7$ case with the binding energy of $^{12}$C shifted to be identical to that obtained with $10^8$ states. While the repulsive phase shifts are almost identical in all cases, there is a clear shift in resonance positions when comparing the dashed and solid lines, which is corrected by using the same (RGM-B) value of the $^{12}$C ground state energy in both calculation, see the text for further details.}
\label{fig:NCSMCcomparison}
\end{figure}

Next, we make a first comparison of CT-NCSMC phase shifts using CT-NCSM/RGM norm and Hamiltonian kernels computed with $^{12}$C CT-NCSM wave functions having $N_\mathrm{SD}$=5$\times$10$^7$ and $N_\mathrm{SD}$=$10^8$ (Fig.~\ref{fig:NCSMCcomparison}, dashed and solid lines respectively), denoting each of these cases RGM-A and RGM-B, respectively. 
The CT-NCSMC kernels are computed using the same $^{13}$C wave functions in both cases, obtained on bases of $N_\mathrm{SD}=10^8$ states for both the negative and positive parity spectrum. 
While the overall characteristics of the phase shifts, such as the number of resonances in each partial wave and their approximate position, are consistent between the two CT-NCSMC calculations at different $N_\mathrm{SD}$, there is some disagreement in the exact resonance positions. 

The explanation for this disagreement can be traced back to the difference in the neutron threshold which for all single-nucleon-projectile cases can be identified by the binding energy of the target nucleus. 
By using the same (RGM-B) value of the $^{12}$C ground state energy in both calculations, the phase shifts become essentially identical in both the RGM-A and RGM-B (compare solid lines and points in Fig.~\ref{fig:NCSMCcomparison}). 

Far from unexpected, this result reaffirms that the projectile-target interaction potential (RGM kernels and NCSMC couplings) arise mostly from the bulk of the wave function, meaning that one can compute them with the low-fidelity wave functions and combine them with independently calculated binding energies that in turn determine decay thresholds. 
There are multiple choices for obtaining the thresholds, either by performing high-fidelity calculations, extrapolating to the infinite-basis result~\cite{Furnstahl2012,Maris2014} or, when precise comparisons to experiment are the goal, setting them to the measured values. 
This latter approach is already in use in the so-called NCSMC-pheno approach~\cite{Hupin2019,Hebborn2022,Kravvaris2023}. 
% Since the energies obtained with tunable-fidelity wave functions follow the same convergence pattern as the $N_\mathrm{max}$ truncation, we can further obtain multiple points to constrain the extrapolation function~\cite{Furnstahl2012,Maris2014} resulting in more confident predictions.  
% Finally, it should be noted that it is reasonable to consider constructing a basis for $^{13}$C in a manner that will treat the thresholds somewhat consistently, i.e. starting from the configurations selected for $^{12}$C and attaching a single neutron to each one, thus ensuring that an underlying connection between the two bases exists. 
% Clearly, such an approach would be less versatile and would probably result in a larger amount of SDs selected for $^{13}$C, but it is not further explored in this work. 

\begin{figure}[t]
\centering
\includegraphics[width=0.47\textwidth]{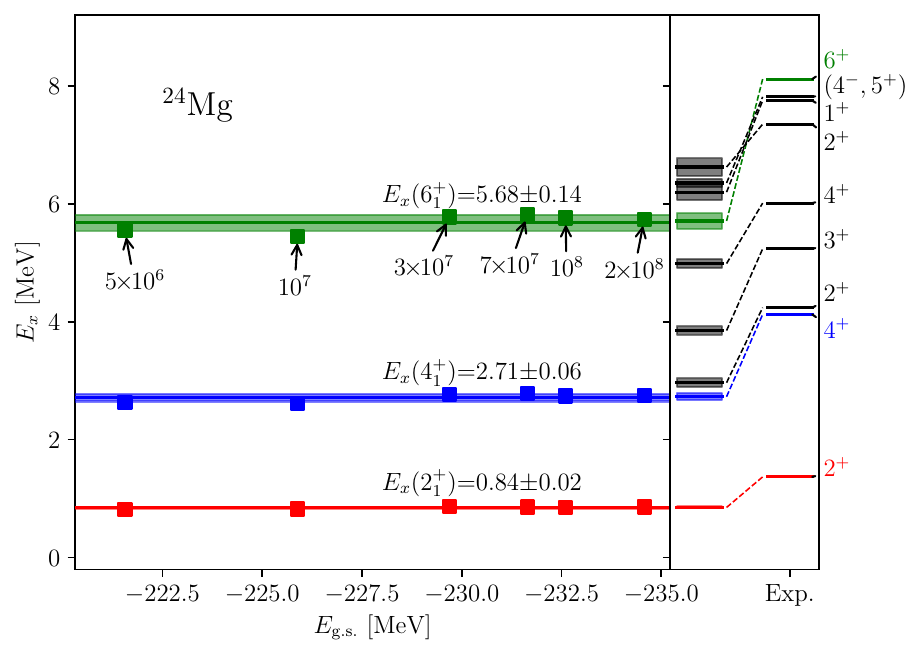} 
\caption{Convergence of excitation energies of the yrast rotational band members for $^{24}$Mg in the CT-NCSM, plotted against the ground state energy ($E_\mathrm{g.s.}$) obtained in the same calculation. While $E_\mathrm{g.s.}$ varies by about 5\% in basis sizes ranging from 5$\times$10$^6$ to 2$\times$10$^8$, the excitation energies remain remarkably constant. Right panel compares the low-lying positive parity states obtained in the CT-NCSM with experimental data. Spin-parity assignments from ENSDF database as of April 1, 2022; the tentative (4$^-$,5$^+$) level is identified as 5$^+$ in the CT-NCSM calculation. The CT-NCSM excitation energies are obtained by extrapolation, with the resulting extrapolation uncertainty denoted by the width of each level.}
\label{fig:Mg24cvg}
\end{figure}

\subsection{Heavier systems}

Finally, we discuss the applicability of the CT-NCSM to nuclei with mass number greater than $A=13$ and specifically to $sd$-shell nuclei. While the IT-NCSM has been applied to nuclei reaching as far as $^{26}$O~\cite{Hergert2013} and even $^{40}$Ca~\cite{Roth2007}, the IT-NCSMC has thus far been applied to lighter nuclei, reaching $^{16}$O~\cite{Navratil2010} and $^{17}$C~\cite{Smalley2015}. 
To this end, we performed exploratory CT-NCSM/RGM calculations of the $n+^{24}$Mg system, lying squarely out of reach of the traditional NCSM/RGM. 
The convergence of the low-spin members of the yrast band computed in the CT-NCSM is seen in Figure~\ref{fig:Mg24cvg}, with the excitation energies ($E_x$)pointing to an almost perfect rotor, with an $R_{42}=E_x(4^+)/E_x(2^+)=3.2$ and $R_{64}=E_x(6^+)/E_x(4^+)=2.1$ (compare with perfect rotor values of 3.3 and 2.1, respectively). 
The nucleus is significantly overbound due to the choice of retaining only the NN part of the SRG-evolved interaction~\cite{Navratil2010}, and the rotational band appears to be somewhat squeezed, suggesting a larger moment of inertia than seen in experiment.  
Nevertheless, the entirety of the low-lying positive-parity spectrum is mostly reproduced (Figure~\ref{fig:Mg24cvg}), yielding evidence for the tentative (4$^-$, 5$^+$) state reported in ENSDF to correspond to the CT-NCSM-computed 5$^+_1$, since no other such spin-parity state is seen up to 11 MeV of excitation. For a more robust spin-parity assignment, the negative parity spectrum should also be computed in a consistent calculation, but this is not done in this work.
Future work including the 3N components of the interaction in the CT-NCSM should provide a clearer picture of the static properties of the nucleus, as well as comparisons with interactions that include, for example, $\Delta-$isobar degrees of freedom~\cite{Novario2020} and can point to how this spectrum arises from the fundamental properties of the nuclear interaction.

\begin{figure}[t]
\centering
\includegraphics[width=0.47\textwidth]{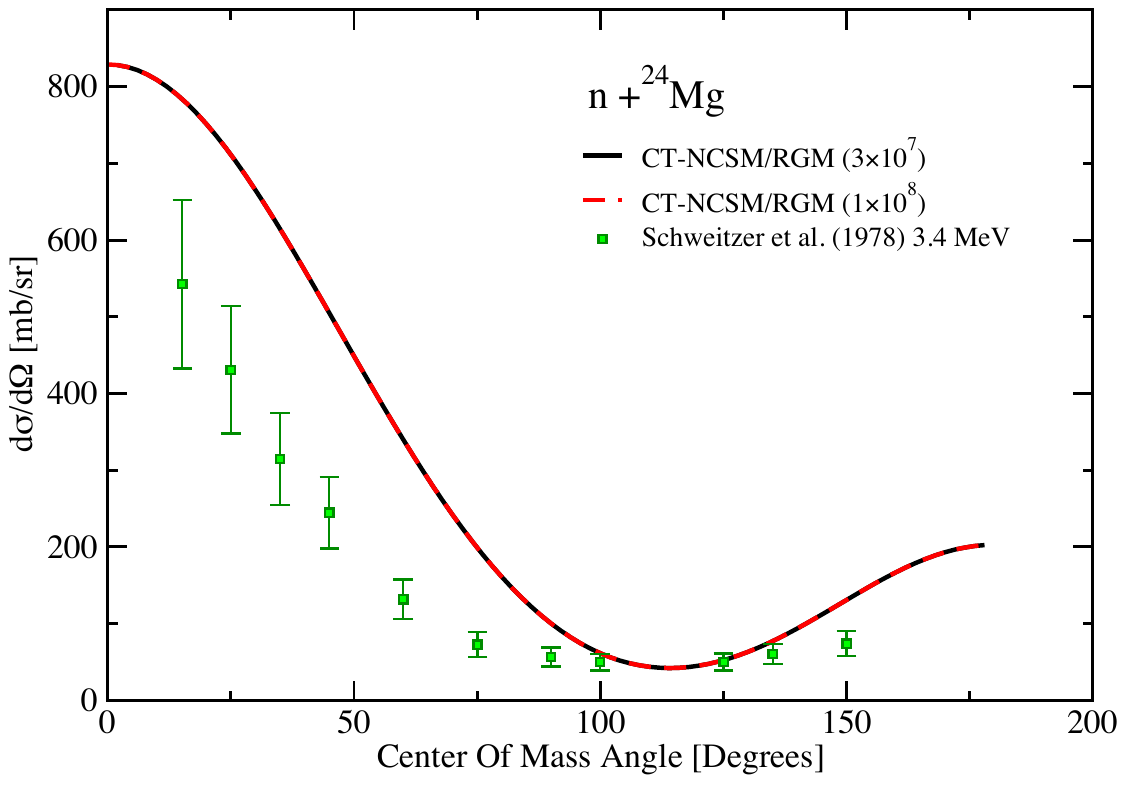} 
\caption{Differential elastic cross section of neutrons scattering on $^{24}$Mg in the CT-NCSM/RGM compared to experimental data from Ref.~\cite{Schweitzer1977}. Tripling the size of the basis used to describe the $^{24}$Mg target ground state has virtually no effect on the cross section.}
\label{fig:Mg24diffxs}
\end{figure}

Despite this overbinding, the excitation energies are well-converged, leading us to the conclusion that the dynamic properties should also converge in a similar manner, much like the $n+^{12}$C case. 
This is indeed the case with, for example, the differential cross section of neutrons with an energy of 3.4 MeV scattering off a $^{24}$Mg nucleus (Figure~\ref{fig:Mg24diffxs}). 
We find a similar convergence when looking at the scattering phase shifts for all angular momentum channels. 
Again, similar to the $n+^{12}$C case, the bound state spectrum in the CT-NCSM/RGM is somewhat less converged, predicting, however, a 1/2$^+$ ground state, a $5/2^+$ first excited state approximately 500 keV higher and a $3/2^+$ state at about 4 MeV of excitation. 
Compared to experiment, the ordering of the first two levels is inverted, and the spacing with the second excited state is significantly larger. 
In contrast, CT-NCSM results for $^{25}$Mg still show a 1/2$^+$ ground state but place the 3/2$^+$ as the first excited state at approximately 300 keV and the 5/2$^+$ at approximately 500 keV. 
Multiple other bound states predicted in the CT-NCSM are not seen in the $n+^{24}$Mg CT-NCSM/RGM calculation as they would arise predominantly from the coupling of excited states of $^{24}$Mg to the incoming neutron which are not included here. 

In future work, the 3N components of the Hamiltonian of Eq.~(\ref{eq:ham}), both SRG-induced and chiral, will be included in the CT-NCSM(C) calculations so that particular effects about the emergence of collectivity in $sd$-shell nuclei can be probed.

%A final question that should be addressed is what the potential limits of applicability of the proposed approach are in terms of nuclear mass. While one could always select a manageable number of SDs for any nucleus, it is absolutely not necessary that the selected basis would provide a good description of the nuclear states. Especially in heavier mid-mass nuclei where the number of states grows significantly faster it is possible that selected manageable basis sizes will not contain high $N_\mathrm{max}$ components. 

Looking further, from the results presented in the previous sections we can estimate that, at a minimum, basis components entering at $N_\mathrm{max}=6-8$ are required for the CT-NCSM  approach to converge. 
At the same time, computational complexity places a limit on how many states can be reasonably included in CT-NCSMC calculations. Being optimistic, we can estimate this upper bound at around 10$^{9}-10^{10}$ basis states. For $^{40}$Ca, states originating from $N_\mathrm{max}=10$ configurations enter after about 2$\times$10$^8$ total SDs, suggesting that reactions involving nuclei across the $sd$-shell can be addressed using the CT-NCSMC. 
We did not pursue CT-NCSM calculations for $fp$-shell nuclei in this work due to computational limitations, but it seems very likely that the lower $fp$-shell would be accessible with minor code improvements in computer memory management.
Preliminary calculations for $^{40}$Ca have shown that the centroids of configurations belonging to different total numbers of quanta ($N$) are significantly more mixed, but still maintain the character seen for $^{12}$C in Figure~\ref{fig:centroids12C}, making the CT-NCSM a viable approach to obtain nuclear wave functions for use in the CT-NCSM/RGM and CT-NCSMC.

It is worth noting, however, that it is not immediately obvious that the configuration selection performed strictly on the basis of the centroid energy is the best choice. 
Indeed, once the configurations that belong to the same $N_\mathrm{max}$ value are sorted according to their centroid energy, we could select the lowest ones per $N_\mathrm{max}$ instead of in an absolute scale as done in the present work. 
Such an approach would ensure that ``long-range" orbits are included, while still keeping the size of the basis manageable; we defer this investigation to future work, along with various other options for selecting configurations including higher Hamiltonian moments.

\section{Conclusions}
In this work we applied the configuration centroid truncation method proposed in Ref.~\cite{Horoi1994} to the NCSM, as well as its NCSM/RGM and NCSMC extensions that allow for the \textit{ab initio} treatment of nuclear reactions. 
Despite the simplicity of the approach, extending from the traditional shell model to the NCSM with its vastly increased number of configurations represented somewhat of a technical challenge. 
We found promising results for both the convergence of energy levels but especially for calculations of reaction observables. 
Specifically, for the case of neutrons scattering off of $^{12}$C in the CT-NCSM/RGM framework, we find rapid convergence, even with modest basis sizes, lending credence to the postulate that the projectile-target interaction does not depend on the more delicate features of the target states, but rather the longer-range properties of the nuclear wave function. 
The precise description of well-bound states,  however, still requires the inclusion of short range correlations in the many-body wave function and thus necessitates the extension to the CT-NCSMC.

When more than one nuclear wave function is needed for the description of the reaction--as is the case for the NCSMC and, in general, for any reaction except for single-nucleon elastic scattering--we find that the position of resonances depends the computed position of the respective threshold which is less accurately described solely within the presented framework. 
Nevertheless, we have pointed to various approaches for mitigating this issue, each with its own set of advantages. Particularly interesting when comparing to low-energy experimental data, the NCSMC-pheno approach described in~\cite{Hupin2015,Hebborn2022,Kravvaris2023} tunes the position of thresholds and resonance positions to the experimental values.

The results presented here enable several avenues of future research. \textit{Ab initio} calculations of nuclear reactions for nuclei up to $^{40}$Ca now appear possible, allowing for a full study of continuum effects along entire isotopic chains, including neutron-heavy light nuclei, where the open-quantum-system effects can be significant, essentially defining the dripline~\cite{Volya2006}. Moreover, as evidenced by the convergence of the $^7$Li spectrum, essential features of deformed nuclei are also well-captured within this approach, making the study of the effects of deformation on reaction cross sections from first principles another possibility for the future. 

This work is the first in an on-going program that aims to both extend the reach of current \textit{ab initio} nuclear reaction methods, as well as connect \textit{ab initio} nuclear theory to phenomenological nuclear reaction methods targeting heavier nuclei relevant in nuclear astrophysics and nuclear technology applications. 

\begin{acknowledgements}
Computing support for this work came from the LLNL institutional Computing Grand Challenge
program.
Prepared by LLNL under Contract DE-AC52-07NA27344. 
This material is based upon work supported by the U.S.\ Department of Energy, Office of Science, Office of Nuclear Physics, 
under Work Proposal No.\ SCW0498. KK receives partial support through the Scientific Discovery through Advanced Computing (SciDAC) program funded by U.S. Department of Energy, Office of Science, Advanced Scientific Computing Research and Nuclear Physics. P.N. acknowledges support from the NSERC Grant No. SAPIN-2022-00019. TRIUMF receives federal funding via a contribution agreement with the National Research Council of Canada.fig
\end{acknowledgements}
%\bibliography{biblio}
%apsrev4-2.bst 2019-01-14 (MD) hand-edited version of apsrev4-1.bst
%Control: key (0)
%Control: author (72) initials jnrlst
%Control: editor formatted (1) identically to author
%Control: production of article title (-1) disabled
%Control: page (0) single
%Control: year (1) truncated
%Control: production of eprint (0) enabled
%
\end{document}